\documentclass[pdflatex,sn-mathphys-ay]{sn-jnl}

\usepackage{graphicx}%
\usepackage{multirow}%
\usepackage{amsmath,amssymb,amsfonts}%
\usepackage{amsthm}%
\usepackage{mathrsfs}%
\usepackage[title]{appendix}%
\usepackage{xcolor}%
\usepackage{textcomp}%
\usepackage{manyfoot}%

\usepackage{booktabs}%
\usepackage{algorithm}%
\usepackage{algorithmicx}%
\newcommand{\MOKA}{\ensuremath{\mathrm{MOKA^{3D}}}}

\def\msun{M$_\odot$}

\def\arcsec{$^{\prime\prime}$}

\usepackage[numbers]{natbib}

\usepackage{algpseudocode}%
\usepackage{listings}%

\begin{document}

\title[Unveiling the Fast Acceleration of AGN-Driven Winds at Kiloparsec Scales]{Unveiling the Fast Acceleration of AGN-Driven Winds at Kiloparsec Scales}

\author*[1,2]{\fnm{Cosimo} \sur{Marconcini}}\email{cosimo.marconcini@unifi.it}

\author[1,2]{\fnm{Alessandro} \sur{Marconi}}

\author[2]{\fnm{Giovanni} \sur{Cresci}}

\author[2]{\fnm{Filippo} \sur{Mannucci}}

\author[1,2,3]{\fnm{Lorenzo} \sur{Ulivi}}

\author[2,4,5]{\fnm{Giacomo} \sur{Venturi}}

\author[1,2,3]{\fnm{Martina} \sur{Scialpi}}

\author[6]{\fnm{Giulia} \sur{Tozzi}}

\author[2]{\fnm{Francesco} \sur{Belfiore}}

\author[2]{\fnm{Elena} \sur{Bertola}}

\author[4]{\fnm{Stefano} \sur{Carniani}}

\author[1,2]{\fnm{Elisa} \sur{Cataldi}}

\author[2]{\fnm{Avinanda} \sur{Chakraborty}}

\author[2]{\fnm{Quirino} \sur{D'Amato}}

\author[1,2]{\fnm{Enrico} \sur{Di Teodoro}}

\author[2]{\fnm{Anna} \sur{Feltre}}

\author[2,6]{\fnm{Michele} \sur{Ginolfi}}

\author[1,2]{\fnm{Bianca} \sur{Moreschini}}

\author[1,2]{\fnm{Nicole} \sur{Orientale}}

\author[1,2]{\fnm{Bartolomeo} \sur{Trefoloni}}

\author[7,8,9]{\fnm{Andrew} \sur{King}}

\affil*[1]{\orgdiv{Dipartimento di Fisica e Astronomia}, \orgname{Università degli Studi di Firenze}, \orgaddress{\street{Via G. Sansone 1}, \city{Firenze}, \postcode{I-50019}, \state{Sesto Fiorentino, Firenze}, \country{Italy}}}

\affil[2]{\orgname{INAF - Osservatorio Astrofisico di Arcetri}, \orgaddress{\street{Largo E. Fermi 5}, \city{Firenze}, \postcode{I-50125}, \country{Italy}}}

\affil[3]{\orgname{University of Trento}, \orgaddress{\street{Via Sommarive 14}, \city{Trento}, \postcode{I-38123}, \country{Italy}}}

\affil[4]{\orgname{Scuola Normale Superiore}, \orgaddress{\street{Piazza dei Cavalieri 7}, \city{Pisa}, \postcode{I-56126}, \country{Italy}}}

\affil[5]{\orgname{Instituto de Astrofísica, Facultad de Física, Pontificia Universidad Católica de Chile}, \orgaddress{\street{Casilla 306}, \city{Santiago}, \postcode{22}, \country{Chile}}}

\affil[6]{\orgname{Max-Planck-Institut für Extraterrestrische Physik (MPE)}, \orgaddress{\street{Giessenbachstraße 1}, \city{Garching}, \postcode{D-85748}, \country{Germany}}}

\affil[7]{\orgname{School of Physics \& Astronomy, University of Leicester}, \orgaddress{\street{}, \city{Leicester}, \postcode{LE1 7RH}, \country{UK}}}

\affil[8]{\orgname{Astronomical Institute Anton Pannekoek, University of Amsterdam}, \orgaddress{\street{Science Park 94}, \city{Amsterdam}, \postcode{NL-1098 XH}, \country{the Netherlands}}}

\affil[9]{\orgname{Leiden Observatory, Leiden University}, \orgaddress{\street{Niels Bohrweg 2}, \city{Leiden}, \postcode{NL-2333 CA}, \country{the Netherlands}}}

\abstract{Supermassive black holes at the centre of galaxies gain mass through accretion disks. Models predict that quasi-spherical winds, expelled by the black hole during active accretion phases, have a key role in shaping galaxy evolution by regulating star formation, the distribution of metals over kiloparsec scales, and by sweeping ambient gas to the outskirts and beyond of galaxies \cite{dimatteo2005, Cattaneo2009, fabian2012, Cicone2018}.
Nonetheless, the mechanism driving these outflows and the amount of energy exchanged between the wind and the galaxy’s interstellar medium (ISM) remain unclear \cite{harrison2018,Wylezalek2018}. Here, we present a detailed analysis of the kinematical properties of winds in a sample of nearby active galaxies using the novel kinematic tool \MOKA, which takes into account the clumpy nature of the ISM. We find remarkable similarities among the properties of the outflows in all the galaxies examined.  In particular, we provide the first evidence that outflows exhibit a regular trend in radial velocity, initially constant or slightly decreasing, followed by rapid acceleration starting at approximately \textbf{1 kpc} from the nucleus, despite the seemingly complex kinematics observed. The observed behavior aligns with our current theoretical understanding of Active Galactic Nuclei (AGN) outflows, where a momentum-driven phase transitions to an energy-conserving phase just beyond approximately 1 kpc \citep{King_pounds_2015}. The constant velocity of the momentum-driven wind is then rapidly accelerated following the inefficient Compton cooling of post-shock material and the transition to energy conservation. Overall, the measured radial terminal velocities of the outflows are always larger than the escape velocities from the host galaxies, confirming the key role of outflows in shaping the galaxy properties and evolution, as a manifestation of AGN feedback.
Our results, only made possible by our novel kinematic analysis tool, are crucial to understand the origin and the powering mechanism of these winds.
}

\maketitle


Multi-wavelength observations of active galactic nuclei (AGN) powered by accretion of matter onto super massive black holes (SMBHs) have revealed the presence of multi-phase winds, observed from pc scale in the X-rays ($v \sim 0.1 c$) to kpc scale for molecular, ionized and atomic outflows ($v \sim 10^{2}-10^{3}$ km s$^{-1}$), at both low and high redshift \citep{Tombesi2010, Sturm2011, Cicone2014, cresci2015a, carniani2015, carniani2016, Cicone2018, harrison2018, fluetsch2019}. 
The exact role and impact of these winds in the context of galaxy evolution, star formation regulation and more in general feedback mechanism is still debated and lack a pivotal observational confirmation. Nonetheless, a plethora of observations and theoretical models suggest that during the most intense active phase of AGN in cosmic history, happening at z $\sim$ 1-3, energetic winds propagating through the galaxy quench star formation by ejecting the gas reservoir out of it and by heating the ISM \citep{perez2008, fabian2012, Zubovas2012, Cicone2014, cresci2015a, carniani2015, cresci2018}.
The outflows powering mechanisms and their energetic impact on the ISM have not been established yet, therefore the  characterization of outflow physical properties is crucial to understand how galaxy evolution processes, such as star formation regulation and gas depletion, are ultimately affected by the AGN feedback mechanism.

Outflows are mostly detected in the ionized phase by the presence of a prominent broad line wings in optical nebular lines, such as [OIII]$\lambda4959,5007$ and [NII]$\lambda6548,6584$, mostly tracing warm (T $\sim 10^{4}$ K) gas ionized by the central AGN \citep{canodiaz2012, Cicone2018, harrison2018}.
Even though nearby, low-luminosity galaxies do not host winds as powerful as those at z $\sim$ 1-3, they represent the best candidates to constrain the energy exchange mechanism between AGN-powered phenomena (outflows and jets) and the ISM. During the last decade, the comprehension of this mechanism has been improved by spatially resolved, integral field unit (IFU) observations, which have proven to be essential to fully understand  outflow properties. In particular, IFU observations of nearby AGN winds allow us to characterize in detail the outflow kinematics and structure from  nuclear up to  galactic scales, taking advantage of the spatial resolution reaching down to a few pc scale for the nearest sources and the high sensitivity that can be achieved \citep[][]{Harrison2014, mingozzi2019, kakkad2023, Speranza2024}.

To constrain the outflow and galactic disc properties, it is pivotal to adopt a kinematic model tailored to account for the complexity of the observed data. Indeed, the impact of the outflow on the host mainly depends on the amount of carried energy, which depends on the outflowing gas mass, velocity and extent, which are  affected by geometrical projection effects and therefore often assumed. The advent of modern ground- and space-based facilities has led to significant improvements in terms of  spatial resolution \citep[e.g.][]{kakkad2023, Chu2024, Lai2022}. Nevertheless, there are still several uncertainties that affect mass and kinematics measurements. 

Most kinematic models rely on assumptions regarding the physical gas properties, and most importantly they are not designed to fully reconstruct the 3D structure of the emitting gas, therefore increasing the uncertainties on the estimated outflow properties \citep[e.g.][]{Crenshaw2000, Das2005, Storchi2010}. As an example, most models assume an analytic smooth gas emissivity which is an extreme simplification of the clumpy and inhomogeneous emission observed in nearby sources. In such a scenario, the observed complexity of gas emission and kinematics cannot be modelled with such simplified models and therefore requires a more comprehensive and accurate analysis.

We apply our novel 3D kinematic model \MOKA, presented in \cite{Marconcini2023} (M23, hereafter), to measure with unprecedented accuracy the kinematics of ionized AGN-driven outflow in a sample of galaxies selected from the MAGNUM (Measuring AGN Under MUSE) survey \citep{cresci2015_magnum, venturi2018, mingozzi2019}. 
Our sample is composed of ten Seyfert galaxies with an inhomogeneous flux distribution in the outflow cones and complex kinematics that so far made difficult to derive reliable kinematical parameters using standard modelling (see Target Selection). Indeed, the complexity of the data, together with the limited capabilities of available kinematic models, could only provide coarse estimates of the outflow velocity, with coarse estimates of the outflow inclination. We take advantage of the capabilities of \MOKA \ to model this complexity and infer the intrinsic outflow velocity profile, providing the first spatially resolved analysis of the outflow geometry and kinematics.  

\begin{figure*}[h!]
\centering
\includegraphics[width=\textwidth]{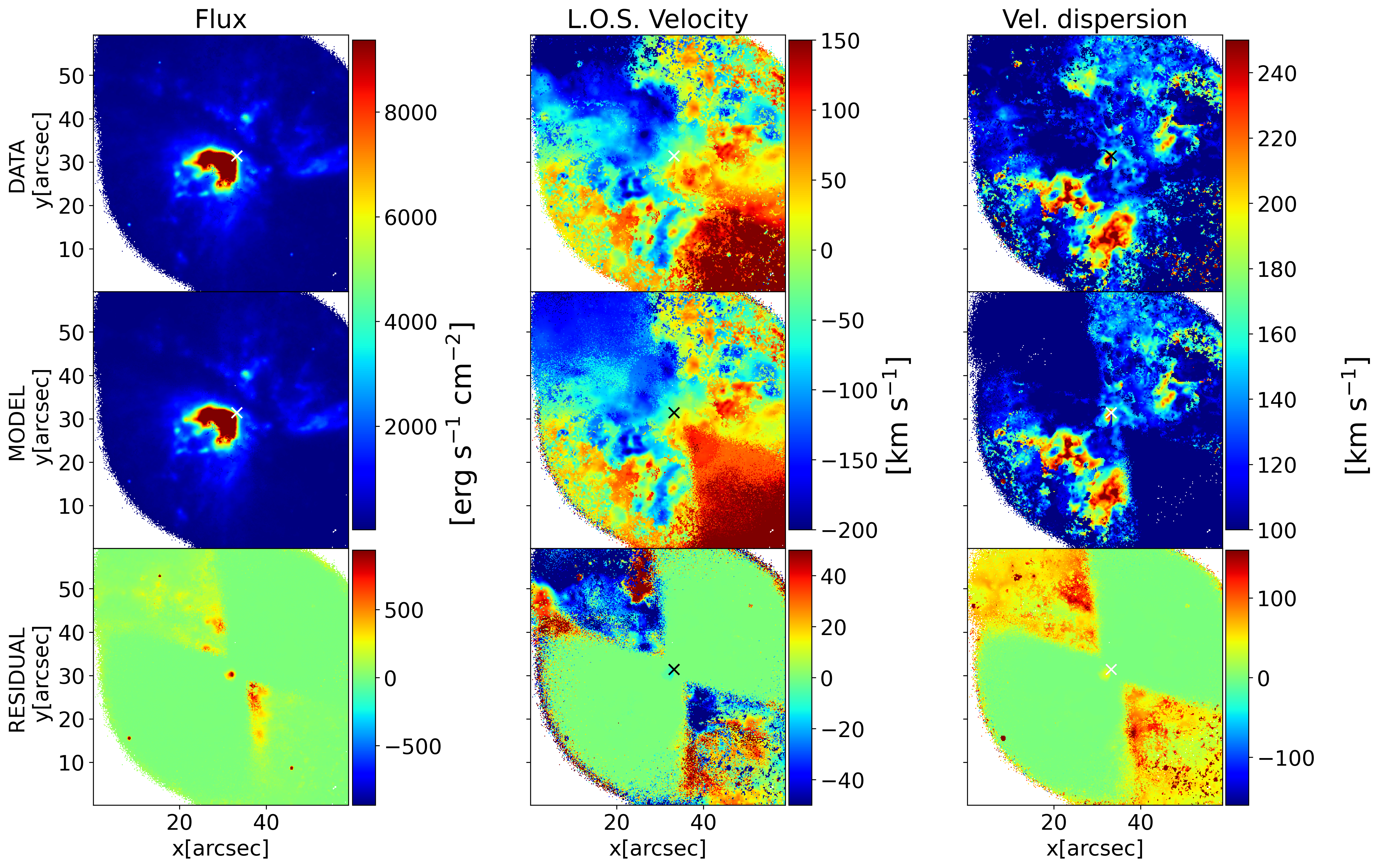}
\caption{Comparison between observed and modelled moment maps for NGC 1365. From top to bottom, panels show the observed, \MOKA \ modelled and residual moment maps. From left to right: integrated flux, line-of-sight velocity and velocity dispersion maps from [OIII]$\lambda$5007 emission (see Methods).}\label{fig:NGC1365_weightedbestfit}
\end{figure*} 


\begin{figure}[htbp]
\centering
\includegraphics[width=\textwidth]{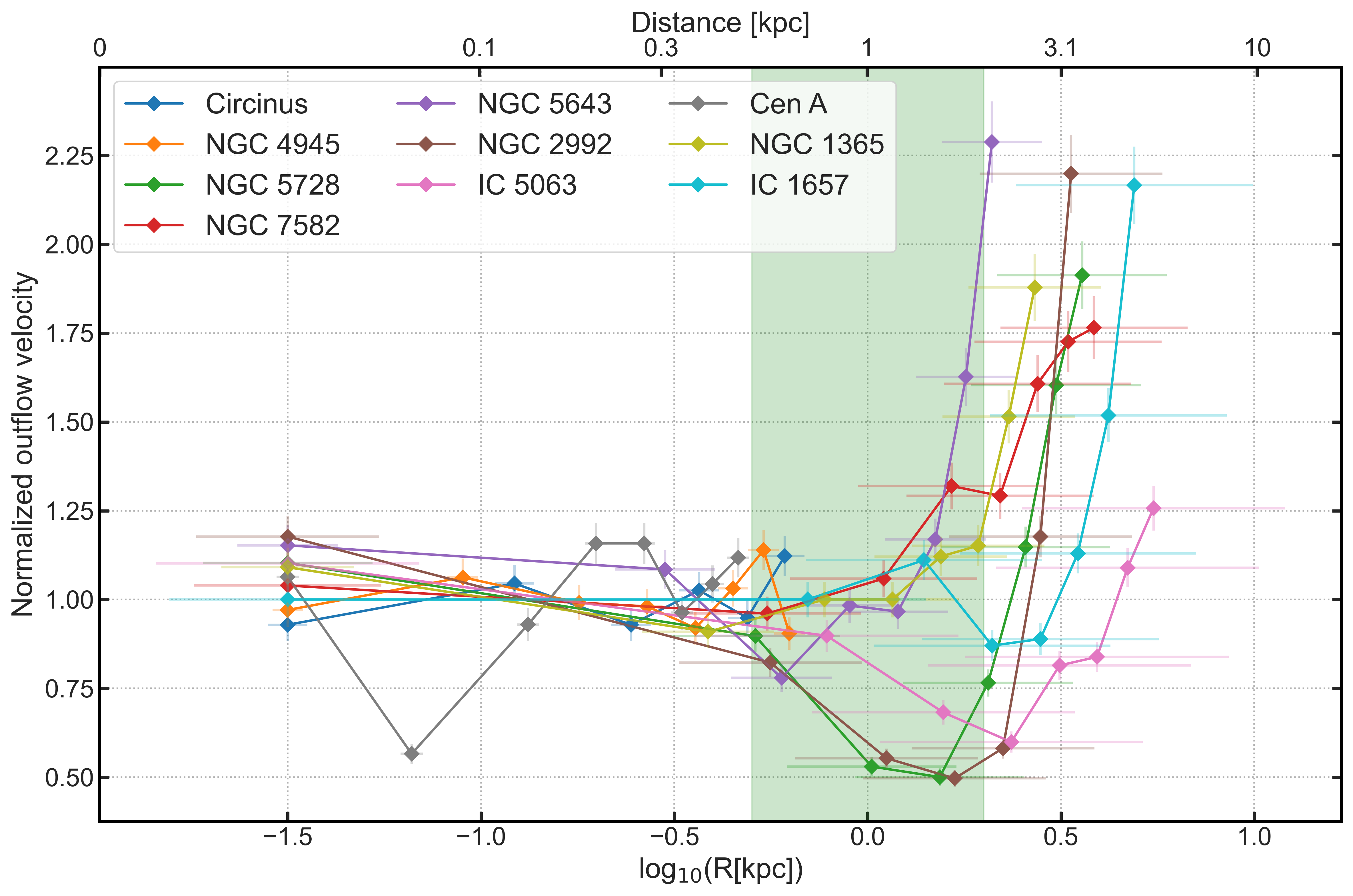}
\caption{Intrinsic outflow radial velocity profile as a function of the distance from the AGN in logarithmic scale. Radial velocity in each shell is inferred with \MOKA. The outflow velocity profiles are normalized to the outflow velocity within 1 kpc from the AGN. The shaded green region marks the putative scale at which theoretical models predict the wind to accelerate due to the transition between momentum- to energy-conserving regimes.}\label{fig:vel_profiles}
\end{figure}

\MOKA \ is a three-dimensional kinematic model that allows us to reproduce the clumpy ionised structures observed in spatially resolved data, finally providing the intrinsic 3D deprojected gas geometry and kinematics (see M23 and Methods for a more detailed description). In order to remove the degeneracies which are present when deriving 3D structures from the observed 2D projections on sky, we minimized the numbers of free parameters by considering conical outflows with pure radial velocities: the cones where divided into concentric conical shells and free parameters were, for each shell, only the outflow velocity ($\rm V_{out}$) and inclination with respect the line of sight ($\beta$). 

As an example of the model capabilities, we discuss the results of the modelling for the nearby Seyfert-II galaxy NGC 1365, which has already been extensively studied with standard techniques \citep{venturi2018}. Figure \ref{fig:NGC1365_weightedbestfit} shows from top to bottom the observed, modelled and residual moment maps and from left to right the integrated [OIII]$\lambda$5007 emission, line-of-sight velocity and velocity dispersion maps. 
With \MOKA \ we retrieve the outflow and disk intrinsic (i.e. de-projected) velocity profiles as a function of the distance from the AGN. Figure \ref{fig:NGC1365_weightedbestfit} shows that we can reproduce with unprecedented accuracy the clumpy ionised outflow emission with a conical geometry and a radial velocity field combined with a rotating disc to model the gas emission in the galaxy disc (the same applies to the rest of the sample, see Sec. \ref{appendix2_moka_fit}). 
For this galaxy, we found that the outflow geometry is well reproduced by dispersing clouds in a cone with a total aperture of 120$^{\circ}$ and average inclination with respect to the line of sight (LOS) of 81$^{\circ}$ $\pm$10. The observed ionised emission is explained with constant radial motions within each conical shell. For this specific galaxy, we divided the outflow in 8 concentric shells of fixed width of 3.8\arcsec. We derived intrinsic radial velocities from $\sim$ 900 km~s$^{-1}$ in the inner shell up to 1500 km~s$^{-1}$ in the outer shell, spanning distances from 0.4 to 2.7 kpc from the AGN. 
As a comparison, \cite{venturi2018} analyzed the ionised gas kinematics in the same source, assuming as outflow velocity the peak of the broad component used to fit the line profile. With this assumption and not accounting for projection effects, they found a decelerating outflow velocity profile. In particular, with their assumption they do not account for the broadening of the velocity profile at larger distances, which crucially determine the higher intrinsic outflow velocity. Instead, taking into account projection effects and considering that the outflow physical velocity extends beyond the peak of the broad component (i.e., reaching the 1\% and 99\% percentile of the line flux), our best-fit model reveals an accelerating profile up to the maximum outflow extension, confirming the effectiveness of our modelling in recovering the intrinsic properties of the outflow. 
Finally, Figure \ref{fig:NGC1365_weightedbestfit} also shows that, despite the observed complex kinematics, the intrinsic velocity field can be explained as a simple radial flow, with the complexity mostly due to the clumpy distribution of ionised clouds (see also M23).

Figure \ref{fig:vel_profiles} shows the derived radial velocity profiles for all galaxies in the sample (see Target selection), normalized to the average outflow velocity within 1~kpc. All galaxies are characterized by a constant or slightly decreasing velocity up to $\sim 1$~kpc, after which we observe a rapid acceleration in which the outflow radial velocity almost doubles within 3-5~kpc. We note that the present sample spans a wide range of distances (from $\sim$3 to 50 Mpc) and therefore the similarities in the radial velocity profiles are not related to the apparent angle projected on the sky.
Indeed, in Circinus, NGC 4945 and Cen A, i.e. the nearest sources, the observed field of view is not wide enough to enconpass the 1~kpc scale. In these three nearby galaxies (D $\leq$ 4 Mpc) we measured a constant average outflow velocity up to their maximum extension, consistently with the other targets. Considering the results for the total sample, the radial velocity trend suggests that at an average distance from the central engine of $\sim$ 1 kpc, the outflow accelerates leading to an increase of the radial velocity up to a factor of $\sim$ 2. 

The simple theoretical model of \cite{King2003} states that an expanding bubble, driven by the nuclear wind, interacts with the host ISM and generates shocks. Then, whether the gas shocked by the approaching wind is able to cool efficiently or not, will determine whether the expanding  bubble is in a momentum- or energy-conserving phase, respectively. Therefore, the shocked gas cooling efficiency has an important role on the velocity of the outflowing gas shells. Assuming the inverse Compton cooling as the most common cooling process, the gas is expected to experience a first isothermal momentum-conserving phase with constant velocity, followed by an acceleration when transitioning to the energy-conserving phase. When the wind reacheas a critical radius of $\sim$ 1 kpc, the cooling time becomes equal to the flow time and the wind cannot cool efficiently anymore: the thermal pressure of the post shock material then accelerates the shell of swept-up gas to much larger velocities (for details of theoretical models see App. \ref{appendix3_theoretical_prediction}). 

Figure \ref{fig:vel_profiles} shows how  observations qualitatively match  the  scenario just described. In particular, the gas entrained into the wind is cooling efficiently up to the acceleration point and is maintaining a constant or slightly decreasing velocity profile by dissipating the wind kinetic energy via radiation. On larger scales, the r$^{-2}$ decrease of the AGN radiation energy density finally makes the cooling time longer than the flow time and the energy which is not radiated away causes a violent acceleration of the wind (see App. \ref{appendix3_theoretical_prediction} for details). Although the physical model is obviously oversimplified, the striking similarities between the observed outflow properties and the theoretical predictions suggests that this model is well suited to describe the physics behind the observed velocity profiles. The shaded green area in Fig. \ref{fig:vel_profiles} marks the predicted acceleration region based on such a scenario, with a critical radius of 1 kpc and assuming uncertainties of $\pm$ 0.3 dex.

\begin{figure*}[t!]
\centering
\includegraphics[width=\textwidth]{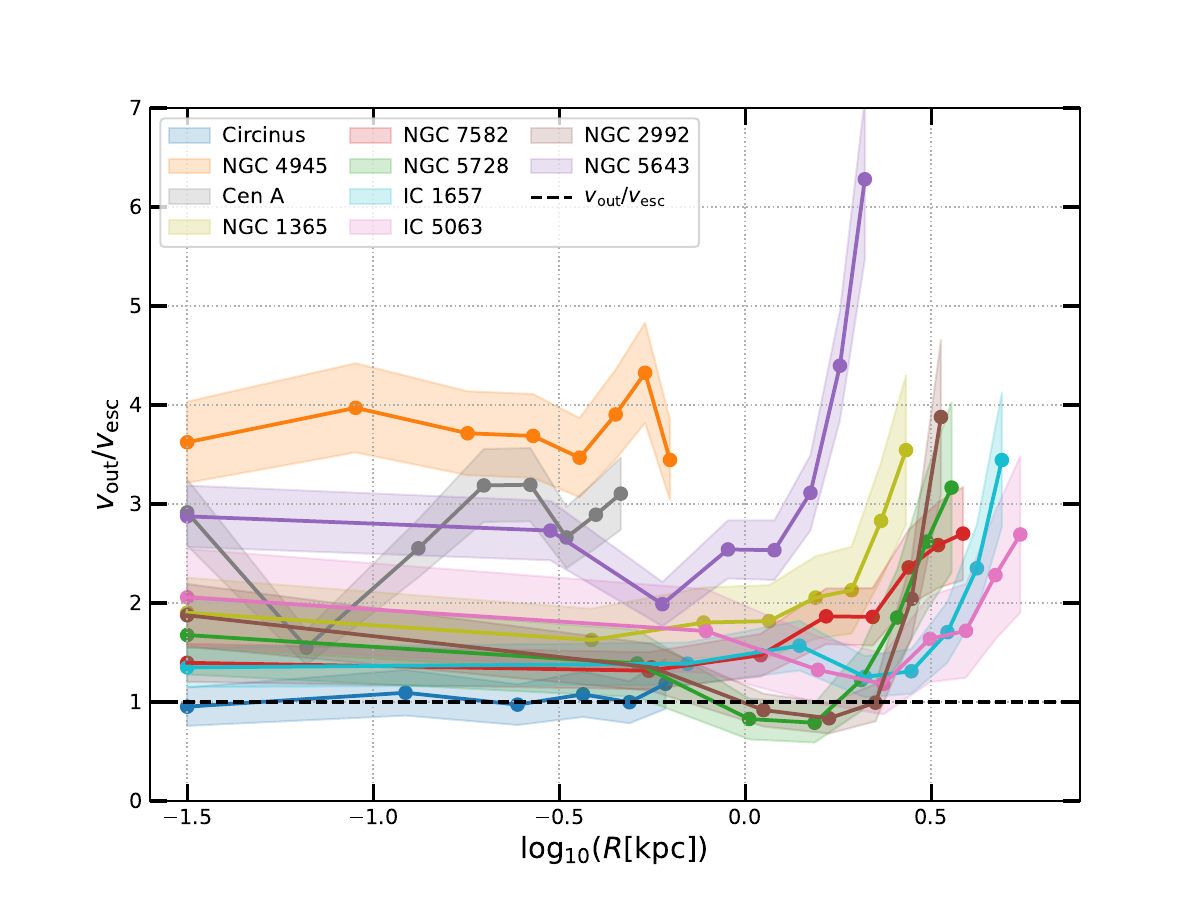}
\caption{Ratio between the outflow intrinsic radial velocity and the host galaxy escape velocity as a function of the distance from the AGN, color-coded by the source name. The dashed black line marks the ratio under which the outflowing gas is bounded by the host potential. The model used to compute the host potential and thus the escape velocity at different radii is discussed in Sec. \ref{escape_velocity}. Shaded areas represent the interval of the ratio computed to estimate the velocity for a particle to reach the a maximum radius between 20 and 100 kpc.}
\label{fig:vout_vesc_profile}
\end{figure*}

Thanks to \MOKA \ we can also fit and de-project the rotational velocity profile of the ionised gas in the galaxy disk and thus, following the prescription in \cite{veilleux2020}, we compute the minimum velocity for the outflowing gas to escape the host galaxy (see Sec. \ref{escape_velocity} for a detailed discussion of the modelling of the galaxy potential using two different methodologies).
Figure \ref{fig:vout_vesc_profile} shows the ratio between the outflow and the escape velocities as a function of the distance from the nucleus. Shaded areas represent the ratio between the outflow and the velocity necessary to escape from the galaxy potential, reaching a 20 or 100 kpc radius. 

We find that in most sources the outflow radial velocity is always higher, up to four times than the escape velocity (see e.g. NGC 4945 and Cen A) and thus the gas is free to escape at any radius. 
In NGC 7582, IC 1657, and IC 5063 the $V_{\rm out}/V_{\rm esc}$ ratio is slightly larger than unity up to a scale of 1 kpc, and then steeply increases through the acceleration of outer shells. Circinus shows a $V_{\rm out}/V_{\rm esc}$ ratio which is slowly increasing with radius, as the outflow velocity remains on average constant up to the maximum extension, while the escape velocity decreases with the distance from the SMBH. 
Finally, we highlight the two interesting profiles of NGC 5728 and NGC 2992. For these two sources, the smooth deceleration of the outflow on the $\sim$ 1 kpc scale might prevent the gas being expelled, establishing a condition of $V_{\rm out}/V_{\rm esc} \leq 1$ for an extensive physical scale. The subsequent acceleration, presumably a result of the transition from a momentum- to energy-driven phase, translates into a drastic increase of the ratio, making it possible for the gas to escape. The trend in Figure \ref{fig:vout_vesc_profile} suggests that the gas entrained in the outflow can reach the galaxy outskirts, so that AGN-powered galactic winds cause feedback by expelling gas, playing a key role in regulating the SMBH growth and star formation by sweeping the ambient gas out of the host potential.

As previously mentioned, AGN-driven outflows may have a crucial role in shaping galaxy properties and evolution and thus it is mandatory to carry out detailed studies to derive the intrinsic outflow physical properties. In particular, characterising the winds kinematics and entrained mass is the first step to constraining the outflow kinetic power and thus the amount of energy transferred to the galaxy ISM. Comparing the outflow kinetic power to the AGN bolometric luminosity (L$_{\rm bol}$) provides a coupling efficiency factor that eventually determines if the observed outflows have a role in shaping galaxy evolution \citep[][]{Storchi2010, Feruglio2015, Harrison2014}. In particular, energetic winds can drastically influence star formation within the host by perturbing and expelling substantial amounts of gas out of the host galaxy, possibly explaining the observed dearth of massive star-forming galaxies in the local Universe \citep[][]{dimatteo2005}. 

Our results highlight how crucial  it is to achieve high intrinsic  spatial resolution and analyse observations with accurate kinematical models in order to uncover the energy-driven phase of AGN winds. Indeed, the theoretically predicted energy-driven regime is established at distances $\ge$ 1 kpc, at which the intrinsic [OIII] luminosity is much smaller compared to that in the circumnuclear region. This means that if both the momentum- and energy-driven regimes are established in an AGN outflow, the latter is harder to observe compared to the former because of the lower luminosity of the emitting gas at larger distances. In all our targets the non-detection of the fainter and more distant energy-driven regime could lead us to assume (incorrectly) much lower intrinsic outflow velocities. If this situation held at all redshifts, it would mean that only the brightest and slowest innermost shells of the outflow would normally be  detected in high-z AGN. This would lead to underestimates of the intrinsic outflow velocities and thus the $V_{out}/V_{esc}$ ratios, with major consequences for the broad picture of galaxy evolution. 
Thus observational limitations, especially at high-z, might cause one to miss contributions to the feedback mechanism. 
We stress the need to obtain robust estimates of the ejected mass rates to specify the efficiency of this channel, and ultimately better understand the AGN feedback mechanism.

In this study, we have provided the first spatially-resolved observational confirmation of the AGN outflow acceleration regime. Moreover, we demonstrate the importance of spatially resolved observations and a suitable model to interpret them. We see that the inefficiency of post-shock cooling is what leads to acceleration of the gas shell and pushes it out of the host potential \citep{ManzanoKing2019, Smethurst2021}. Our results for the behaviour of AGN winds conform closely to theoretical predictions of expanding bubbles, and represent a major step forward in understanding phenomena such as AGN feedback and the mechanism powering it, and the co-evolution of SMBH and their hosts.  In addition they improve our understanding of the impact of outflows on the ISM. In future work, we will apply our model to several other low- and high-z galaxies, adding statistics to our results.

\section{Methods}\label{sec_methods}
\subsection{Observations and spectroscopic analysis}\label{observations_spectroscopic_analysis}
The galaxies in our sample were observed by the optical integral field spectrograph Multi Unit Spectroscopic Explorer (MUSE, \cite{Bacon2010}) at the Very Large Telescope (VLT) in Wide Field Mode (WFM) configuration. The data consist of data cubes with two spatial and one spectral extension.
We analysed each data cube by means of a set of custom Python scripts to first subtract the stellar continuum, and then fit multiple Gaussian components to the emission lines, thus finally obtaining an emission-line model cube for each emission line of interest. For a more detailed description of the data reduction and the spectroscopic analysis we refer to \cite{mingozzi2019, marasco2020, tozzi2021}.
To map outflow properties, we used the [OIII]$\lambda$5007 and [NII]$\lambda$6584 emission lines. They are both optimal tracers of the ionised phase of outflows on $10^{2}-10^{3}$ pc, as they can only be produced in low-density regions and therefore do not trace the sub-parsec scales of the broad line region (BLR).

We created moment maps of the ionised emission from the cube fit: the integrated line flux (moment of order 0), the flux-weighted LOS velocity (moment of order 1) and the velocity dispersion maps (moment of order 2). See top panels in Fig. \ref{fig:NGC1365_weightedbestfit} and Fig. \ref{fig:weighted_moment_maps_cont1} for the observed moment maps of our sample. These maps are fundamental to provide a first hint of the ionised gas kinematics and morphology, to be used as starting guess for the modelling with \MOKA. As stressed in M23, any kinematic model able to reproduce the observed moment maps does not guarantee that the derived properties are a faithful representation of the outflow features and physical properties, due to the high level of degeneracy affecting this kind of analysis, and thus a more accurate line profile modelling is needed (see Sec. 2.6 in M23 for a detailed discussion on outflow kinematic degeneracies).

\subsection{Target selection}\label{appendix1_sample}
Our sample is composed of ten nearby AGN from the MAGNUM survey. These sources are selected by cross-matching the optically selected AGN samples of \cite{maiolino1995} and \cite{risaliti1999}, and Swift-BAT 70-month Hard X-ray Survey \citep{baumgartner2013}, choosing only sources observable from Paranal Observatory (-70$^{\circ} < \delta < 20^{\circ}$) and with a luminosity distance $\rm D_L \leqslant$ 50 Mpc. Distance and SMBH mass of the galaxies in our sample are listed in Tab. \ref{tab.appendix_gal_properties}. In total, the MAGNUM sample collects $\sim$ 80 sources, of which we selected those showing a well-shaped (bi)conical outflow emission traced by [OIII]$\lambda$5007 or [NII]$\lambda$6584 emission lines. Moreover, we decided to select sources spanning the widest possible range of distances, i.e. from NGC 4945 ($\sim$ 3.7 Mpc) to IC 1657 ($\sim$ 50 Mpc) with different scale outflows, from $\sim$ 500 pc in Cen A and NGC 4945 up to $\sim$ 5 kpc in IC 5063. As a consequence of their vicinity and of the high MUSE spatial resolution, all sources show clumpy ionised gas distribution, which is not a selection criterion. In order of distance, the sample is composed by: NGC 4945, Centaurus A, Circinus, NGC 5643, NGC 1365, NGC 7582, NGC 2992, NGC 5728, IC 5063, IC 1657 \citep{cresci2015_magnum, venturi2018, mingozzi2019, Marconcini2023, Zanchettin2023}. 

\subsection{Data modelling with \texorpdfstring{\MOKA}{MOKA}}\label{moka_section}

\begin{figure*}[h!]
\centering
\includegraphics[width=\textwidth]{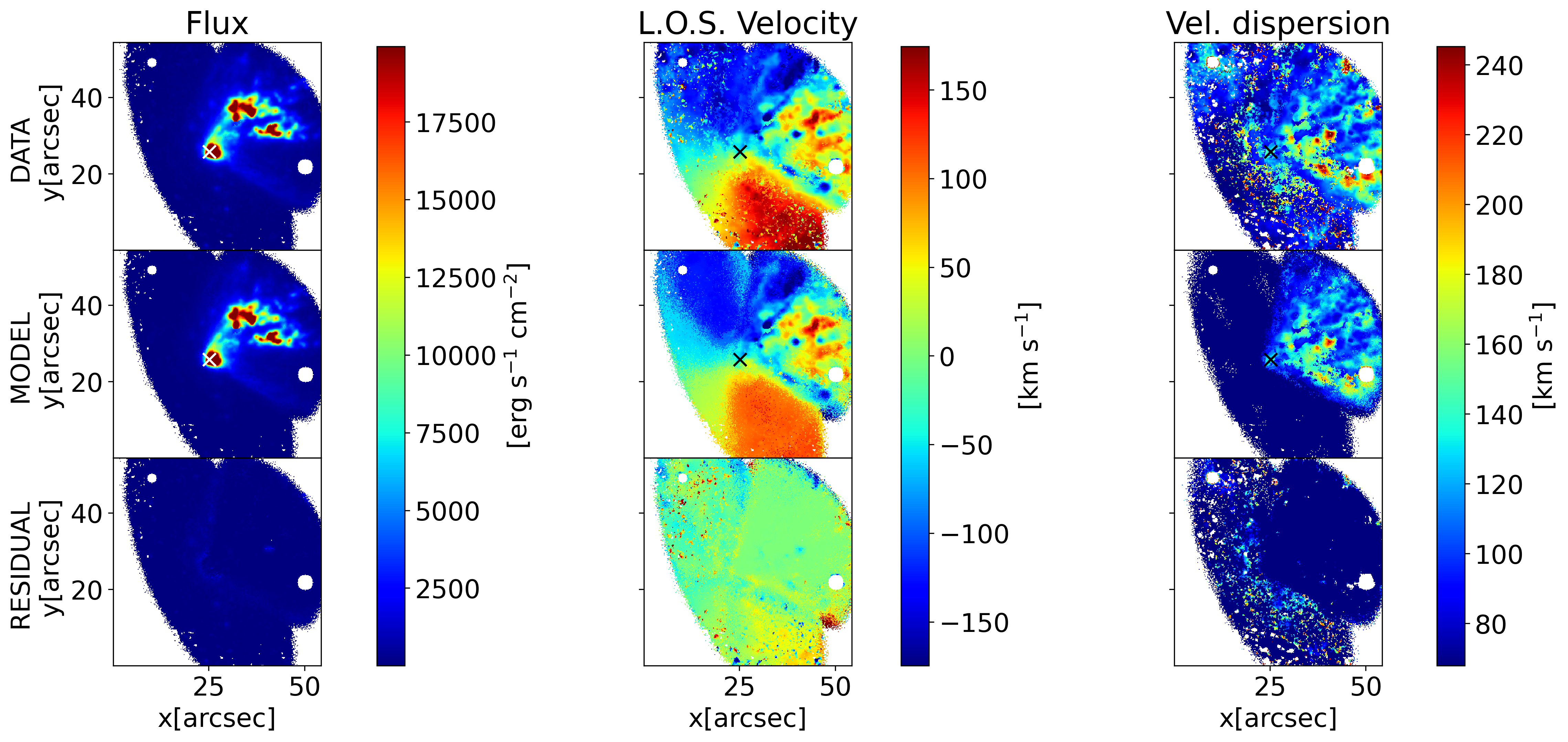}
\includegraphics[width=\textwidth]{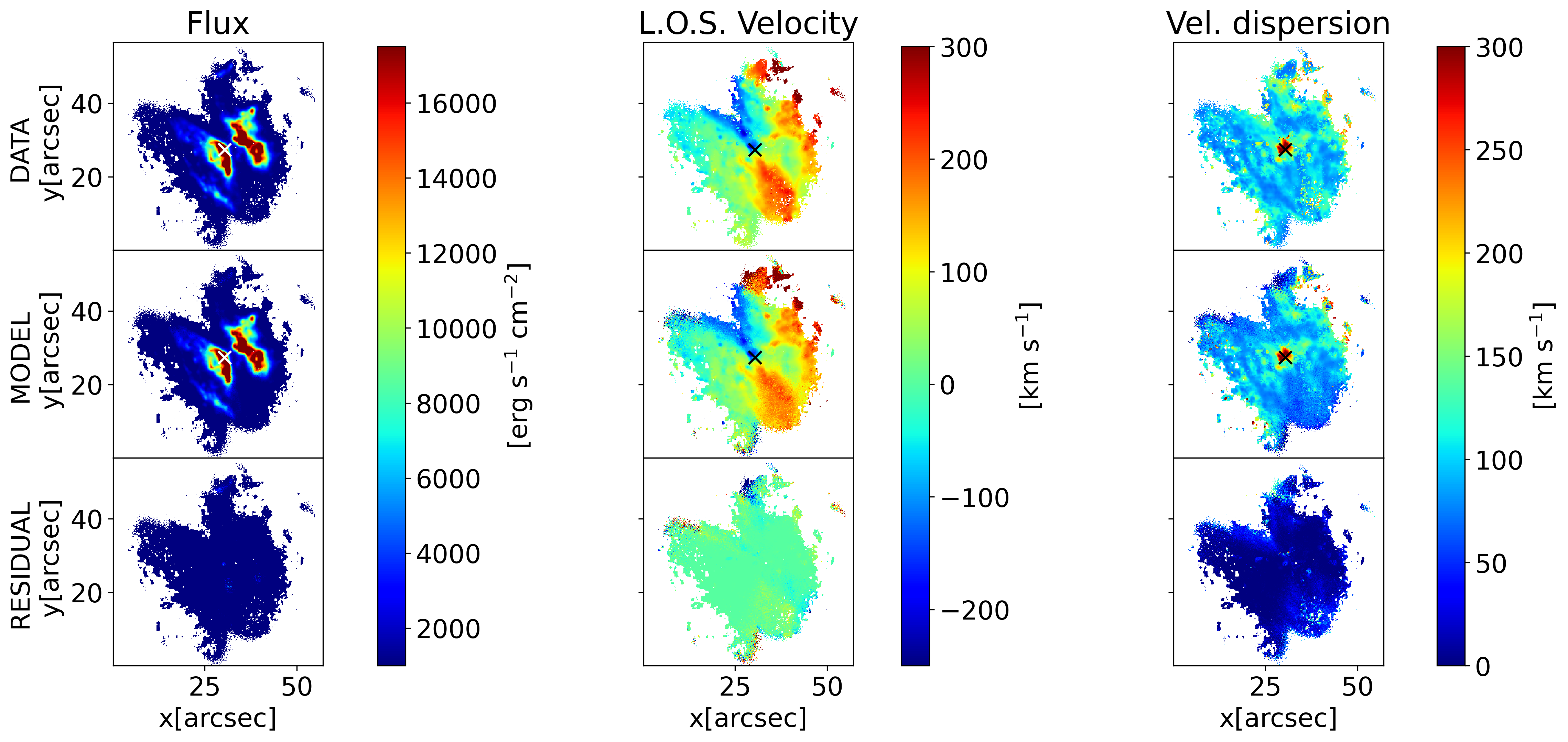}
\includegraphics[width=\textwidth]{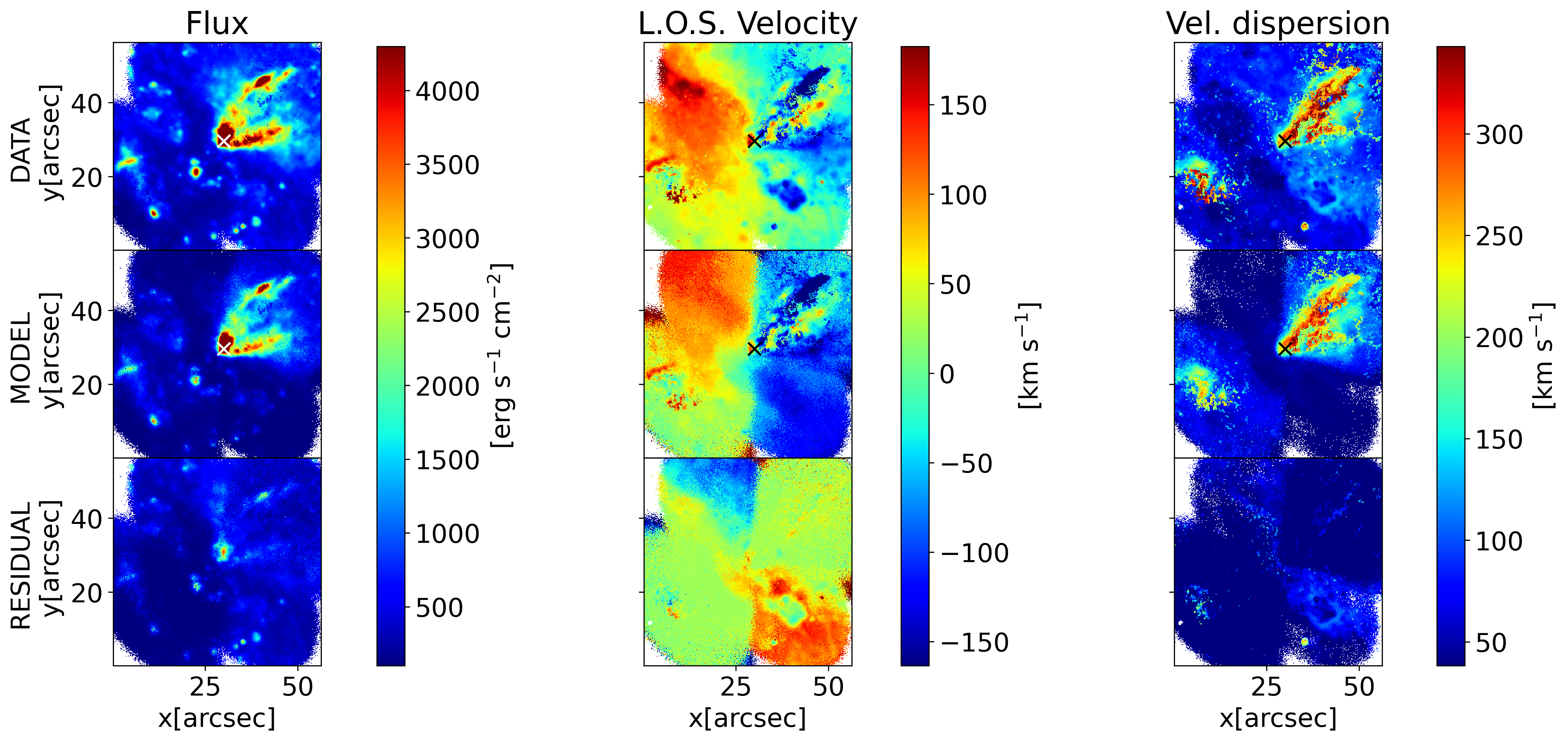}
\caption{Same as Fig. \ref{fig:NGC1365_weightedbestfit} for Circinus, NGC 2992 and NGC 4945. Comparison between \MOKA \ best fit moment maps and masked observations. From top to bottom, panels show the observed, \MOKA \ model and residual moment maps. From left to right: integrated flux, L.O.S. velocity and velocity dispersion maps from [OIII]$\lambda$5007 emission.}\label{fig:weighted_moment_maps_cont1}
\end{figure*}

To model and interpret the kinematics of line emission we adopted the innovative tool \MOKA \ presented in M23.
\MOKA \ is a three-dimensional kinematic model that assumes a spherical or cylindrical geometry populated with a 3D distribution of emitting clouds. These clouds are weighted according to the observed line flux in each spaxel and spectral channel of the data cube. The outflow clouds follow a simple radial velocity field which is free to vary with radius and is based on the chosen geometry and the cloud position in the 3D space. 
Compared to the routine presented in M23, here we improve the tool by implementing a more sophisticated fitting procedure. In particular, in this updated version we reproduce the spatially resolved gas properties as a function of the distance from the nucleus. Some sources in our sample show that the outflow morphology is difficult to constrain due to its superposition with the galaxy disc in projection (e.g. NGC 5643, IC 1657, NGC 1365). Therefore, we combined two kinematic components (outflow + disc) to model the total emission and best reproduce the observed features. In particular, we modelled the outflow with a bi-conical, spherically symmetric model and the disc as a cylindrically symmetric model. In our routine we fit the gas properties dividing the conical outflow and the disc in concentric shells of fixed width. The number of shells and thus their intrinsic width is tailored to the maximum outflow extent in each source in order to minimise their overlap, once projected on the plane of the sky.  The free model parameters in each shell are two: the outflow (disc) intrinsic radial (rotational) velocity and inclination with respect to the line of sight.  First, we masked the outflow spatial shape in the observed moment maps and fitted the rotating gas disk emission in concentric annuli of fixed width, setting each annulus rotational velocity and inclination as free parameters, finally reproducing the observed ionised emission in the galaxy disk. Then, we created a separate model cube for the outflow, assuming that the model clouds follow a radial motion (see Eq. 3 in M23) and combined it with the disk model cube. As discussed in M23, to fit the outflow emission in each source we fixed the position angle ($\gamma$), the outflow conical aperture, and the centre of the outflow model, being parameters that can be assumed from observations. Finally, we fitted the outflow emission in each shell, with the rotating disk underlying it. To do this, we compared the observed and modelled spectra shell by shell. In particular, we compare the observed and modelled percentile velocities at 1$\%$ and 99$\%$ of the emission line integrated over each conical shell, requiring that they differ by no more than 5$\%$. In each shell the free parameters are the intrinsic velocity and the inclination with respect to the line of sight. As a result, we obtained the velocity and inclination as a function of the distance from the nucleus, for each shell of each model\footnote{Unfortunately, \MOKA \ still lacks a detailed analysis of the free parameter uncertainties, nonetheless we believe a 5\% uncertainty on each free parameter to be realistic, since this uncertainty allows to reproduce the observed percentiles at 1\% and 99\% of the line flux with an accuracy $\geq$ 95\%.} (see App. \ref{appendixbo_outflow_disc_inclination} and Fig. \ref{fig:outflow_disk_inclination}). The averaged properties for the outflow and disk kinematics and geometry are summarized in Table \ref{tab_appendix2_outflow_proeprties}.
In Fig. \ref{fig:weighted_moment_maps_cont1}, we show the observed, modelled and residual moment maps for each source in our sample. 
The intensity maps (top left for each source in Fig. \ref{fig:weighted_moment_maps_cont1}) show a clear bi-conical axis-symmetric geometry in each source except Circinus, where the counter cone is probably obscured by the galactic disc \citep{Elmouttie1998}. In each case except for NGC 5728 and IC 1657, the redshifted cone is more dust obscured than the approaching one, as expected in Seyfert galaxies with axis-symmetric conical winds, as the counter-cone is typically partially or totally hidden by the galaxy disk \citep[see e.g.][]{mingozzi2019}. To fit the emission of the ionised gas in the disc, we adopted a thin disc geometry with a disc width of 1\arcsec \ for each source, without including any intrinsic velocity dispersion for the disc clouds. As a consequence, just a few clouds of the disc model will fall in each spaxel and thus we expect higher residuals from the disc-dominated region compared to the outflow-dominated spaxels. This effect can be ascribed to the assumption of modelling the emission from the disc with a thin disc geometry, which is not accounting for non-circular motions and deviations from the simple disc geometry (as disc bulges or thick clumps which deviate from the thin disk geometry).

With \MOKA \ we reproduce both the clumpy ionised emission and complex observed velocity fields demonstrating that the observed complexity has to be ascribed to the clumpy medium and not to the velocity field, which instead can be modelled with a smooth velocity function.
It is worth noticing that such complexity cannot be reproduced by assuming  analytical flux functions which cannot account for the clumpy ISM morphology \citep[e.g.][]{Crenshaw2000, Das2005, Fischer2010, MullerSanchez2011, Bae2016}.
We observe that the effect of increased velocity dispersion perpendicular to bi-conical outflows in NGC 5643, NGC 5728 and IC 5063 (see \cite{venturi2021} for a tentative explanation) is not well reproduced with \MOKA. These features are tracing an enhancement of the turbulence in the ISM which is possibly due to the interaction with the radio jet, and thus cannot be reproduced by our outflow + disc model. Anyhow, the feature is not affecting our modelling of the outflow properties. It is important to stress that the underlying disc emission is negligible to model and reproduce the intrinsic outflow properties. Indeed, in our MAGNUM sample we measured on average rotating disc velocities in a range from a few tens of km s$^{-1}$ in the inner shells up to $\sim$ 250 km s$^{-1}$ in the outer regions. Such velocities are considerably smaller compared to the intrinsic outflow velocities at each radius and thus will not affect our results. 


\begin{table}
  \centering
  \begin{tabular}{c|cccc}
    \hline
    \hline
        ID  & $D$ & $M_{BH}$ & $M_{*}$ & $M_{gas}$ \\
     & [Mpc] & [log(\msun)]& [log(\msun)] & [log(\msun)]  \\
    \hline
    NGC 4945 & 3.7  &  6.15 & 9.7 & 9.8 \\
    Circinus &  4.2 & 6.54 & 10.9 & 9.9  \\
    NGC 7582 & 22.7 & 7.74 & 10.5 & 9.3 \\
    Centaurus A & 3.8 & 7.77 & 9.8 & 9.4 \\
    NGC 1365 & 18.6  & 7.30 & 10.9 & 10.4  \\
    NGC 2992 & 31.5  & 7.36 & 10.8 & 10.7  \\
    IC 5063 & 45.3  & 8.24 & 11.1 & 9.0  \\
    NGC 5643 & 17.3  & 6.44 & X & X\\
    NGC 5728 & 44.7  & 7.53 & 11.1 & 9.4\\
    IC 1657 & 50.5  & 7.67 & 10.6 & X\\

    \hline
  \end{tabular}
  \caption{Properties of the galaxies in our sample. From left to right: Source ID, distance in Mpc, SMBH mass, stellar mass, and gas mass. The SMBH masses (NGC 4945: \cite{Greenhill1997}, Circinus: \cite{Davis2014}, NGC 7582: \cite{wold2006}, Cen A: \cite{cappellari2009}, NGC 1365: \cite{Caglar2020}, NGC 2992: \cite{Caglar2020}, IC 5063: \cite{Kakkad+22}, NGC 5643: \cite{Goulding2010}, NGC 5728: \cite{durre_mould_2018}, IC 1657: \cite{Kakkad+22}), stellar masses (NGC 4945, NGC 7582, Cen A, NGC 2992, NGC 1365, IC 5063, NGC 5728: \cite{lopezcoba2020},  Circinus: \cite{For2012}, IC 1657: \cite{Koss2021}) and gas masses (NGC 4945: \cite{Dahlem1993}, NGC 7582:\cite{Dahlem2005}, Cen A:\cite{Parkin2012}, NGC 2992:\cite{Lianou2019}, NGC 1365:\cite{Jorsater1995}, IC 5063:\cite{Morganti2015}, NGC 5728: \cite{durre_mould_2018},  Circinus: \cite{For2012}, are taken from the literature. The X indicates that to our knowledge there are no estimates from the literature, therefore a stellar and gas mass of 10$^{10}$ \msun\ and 10$^{10}$ \msun\ are assumed in order to compute the escape velocity, respectively.}\label{tab.appendix_gal_properties}
\end{table}

\subsection{\texorpdfstring{\MOKA}{MOKA} fitting and outflow velocity}\label{appendix2_moka_fit}
Our results for the intrinsic outflow average velocity are generally consistent with previous kinematic analysis (see averaged velocities listed in Tab. \ref{tab_appendix2_outflow_proeprties}).
In particular, we compare the results for Circinus, NGC 4945, and NGC 7582 with our previous analysis in M23, where we fitted the data of each source imposing a constant and average outflow velocity and inclination. Here we demonstrated that a constant outflow velocity is still the optimal fit for Circinus and NGC 4945, up to the observed scales. For NGC 7582 instead, to quantify the improvements with respect to our previous analysis and to demonstrate that an accelerating velocity profile is better suited to reproduce the observed features in NGC 7582, we compared the $\chi^2$ for the new best fit model and the one presented in M23. We found that our new routine improves the fitting of up to a factor of 2, i.e. $\kappa_{acc}$/$\kappa_{const}$ = 0.42, with $\kappa_{acc}$ and $\kappa_{const}$ the $\chi^2$ presented in M23 for the accelerating and constant velocity models, respectively. Finally, taking into account the different inclinations, we find that the average values from this work are consistent with the results in M23. 
Consistently with our averaged result, \cite{Fonsecafaria2021} and \cite{Kakkad+22} studied the same MUSE WFM data of the Circinus galaxy and assumed an outflow expansion velocity of $\sim$ 750 km s$^{-1}$ and 400 km s$^{-1}$, respectively, derived from [OIII] channel maps and from the FWHM of the broad [OIII] component, respectively. In NGC 4945 and NGC 1365, \cite{Venturi2017} found ionised outflow velocities up to 800 km s$^{-1}$ and 500 km s$^{-1}$, respectively, using the $W_{70}$ maps as a proxy of outflow velocity.
In Cen A, \cite{israel2017} found [C~I] and $^{12}$CO outflow velocities below 100 km s$^{-1}$, which is inconsistent with our estimates. \cite{durre_2019} and \cite{Shimizu_2019} traced the outflow velocity in NGC 5728 and found maximum physical values of 400 km s$^{-1}$ and 740 km s$^{-1}$, from Pa$\beta$ and [SiVI] emission, respectively. \cite{Zanchettin2023} confirmed ionised outflow velocities above 1000 km s$^{-1}$ in NGC 2992, consistently with our results and in contrast with previous analysis \citep[e.g.][]{GuoloPereira2021}. In NGC 5643, \cite{garcia_benete_2021} and \cite{cresci2015_magnum} found outflow velocities up to 700 km s$^{-1}$ and 450 km s$^{-1}$, respectively. In the latter, the outflow velocity is measured as the velocity at 10\% of the [OIII] line flux. Taking into account projection effects and considering all the line profile, both these estimates are consistent with our results. To our knowledge, the only estimate of the ionised outflow velocity in IC 1656 is provided by \cite{Kakkad+22}. They used the non-parametric velocity dispersion $w_{80}$ from slit spectra as proxy of outflow velocity and found velocities of $\sim$ 400 km s$^{-1}$, which is below our spatially resolved estimate. Lastly, \cite{Juneau2022} found projected ionised flux-weighted maximum outflow velocities of $\sim$ 300 km s$^{-1}$ in NGC 7582, which is not consistent with our results even de-projecting the outflow velocity. The possible reason for this inconsistency might be that they are observing the flux-weighted gas velocity and thus are not taking into account the contribution from the gas producing the fainter line wings, which would lead to assume a higher outflow velocity.



\subsection{Escape velocity}\label{escape_velocity}
We computed the escape velocity profile for all our MAGNUM galaxies following two different methodologies.
First, based on the prescription of \cite{veilleux2020}, we assumed a spherically symmetric galaxy and modelled the potential as in a singular isothermal truncated sphere (to avoid the mass to diverge we truncated the sphere at R$_{\rm max}$ = R$_{\rm halo}$). Then, to be as conservative as possible, we assumed as disk velocity the maximum disk velocity measured with \MOKA. Finally, we can describe the escape velocity as: $\rm v_{ESC}(r) = V_{rot} \sqrt{2 \left(1 + ln\left(R_{max}/r\right)\right)}$. The second method to estimate the escape velocity profile is based on \citet{Tozzi2024}. We build a dynamical mass model using the python package galpy \cite{bovy2015} and derive the total rotation curves and escape velocity curves. Our mass model consists of a dark matter (DM) halo, a stellar disk, and a gaseous disk. We model the DM halo through a NFW profile, deriving its virial mass $M_{200}$ and concentration $c$ from the stellar-to-halo mass relation of \cite{girelli2020} and the $M_{200}-c$ relation of \cite{dutton2014}, respectively. For the stellar disk, we assume a double-exponential profile with scale-length $R_{\rm d}$ and scale-height $R_{\rm d}$/5, where $R_{\rm d}$ is equal to the half-light radius $R_{50}$ divided by 1.68 (correct for a pure exponential disk, Sérsic index $n = 1$). We determine $R_{50}$ from the size-M$_{*}$ relation for a disk galaxy (n = 1) at z$\sim$ 0 \cite{Mowla2019} adopting the stellar masses listed in Tab. \ref{tab.appendix_gal_properties}. We model a gaseous disk of M$_{\rm gas}$ adopting the values listed in Tab. \ref{tab.appendix_gal_properties}. In doing so, we derive the escape velocity profile of the total mass distribution. Finally, despite the two methods have different assumptions, we derived consistent values for the escape velocity profile. Since both models are well suited to compute the escape velocity in our sample, we decided to adopt the estimates from the first one as it is simpler and does not rely on measurements of galactic properties (e.g. stellar masses, gas mass, SMBH mass) to set the galaxy dynamics.

\begin{figure*}[h!]
\centering
\includegraphics[width=\textwidth]{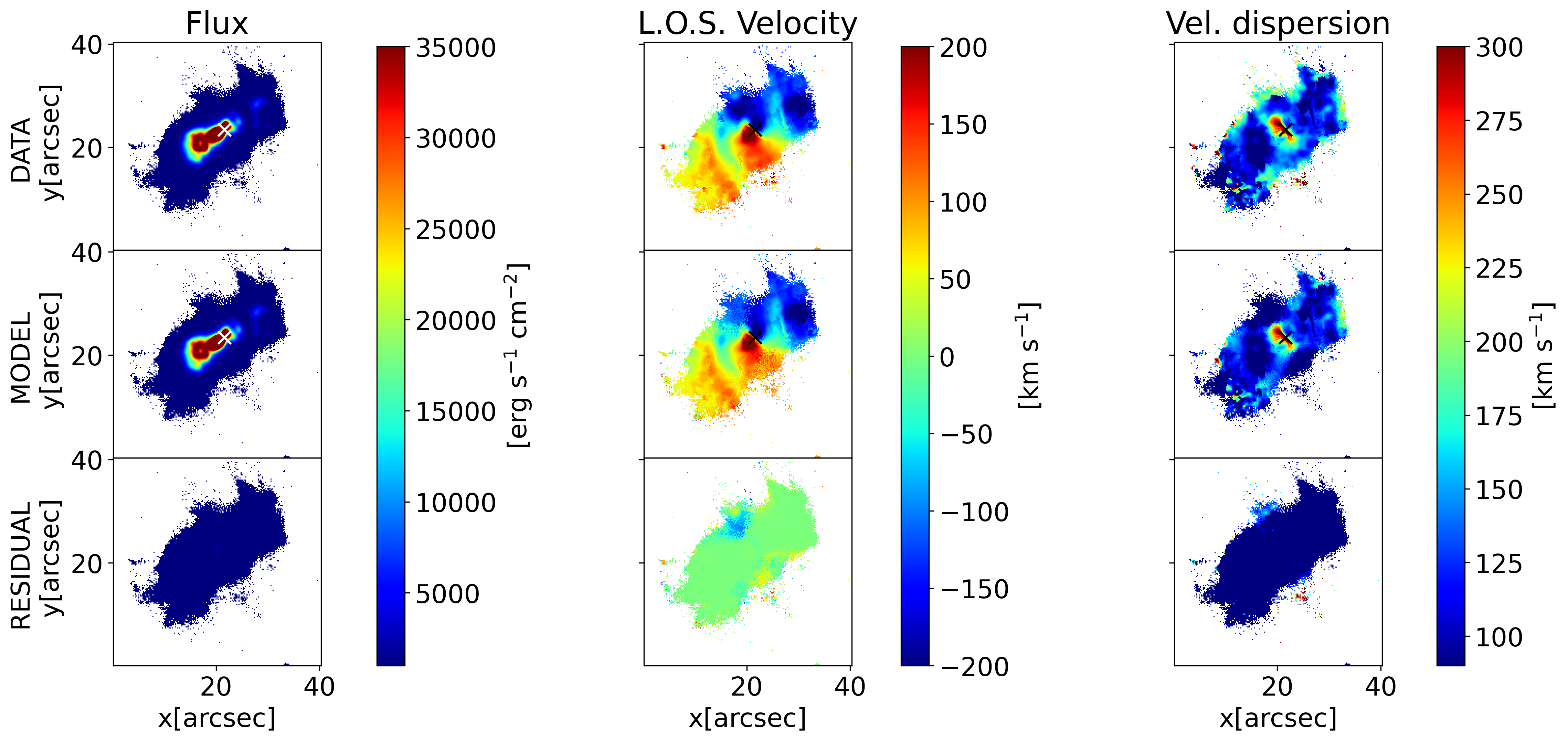}
\includegraphics[width=\textwidth]{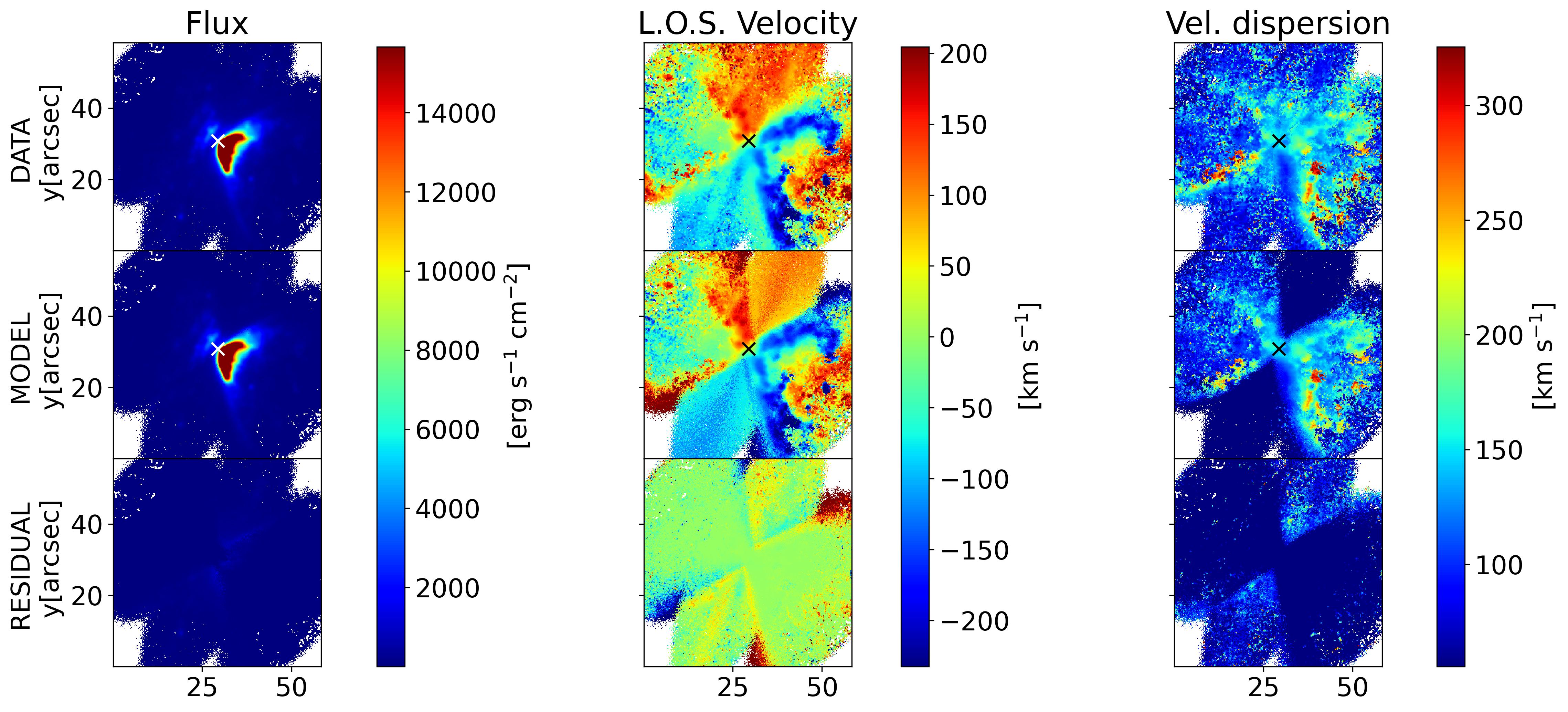}
\includegraphics[width=\textwidth]{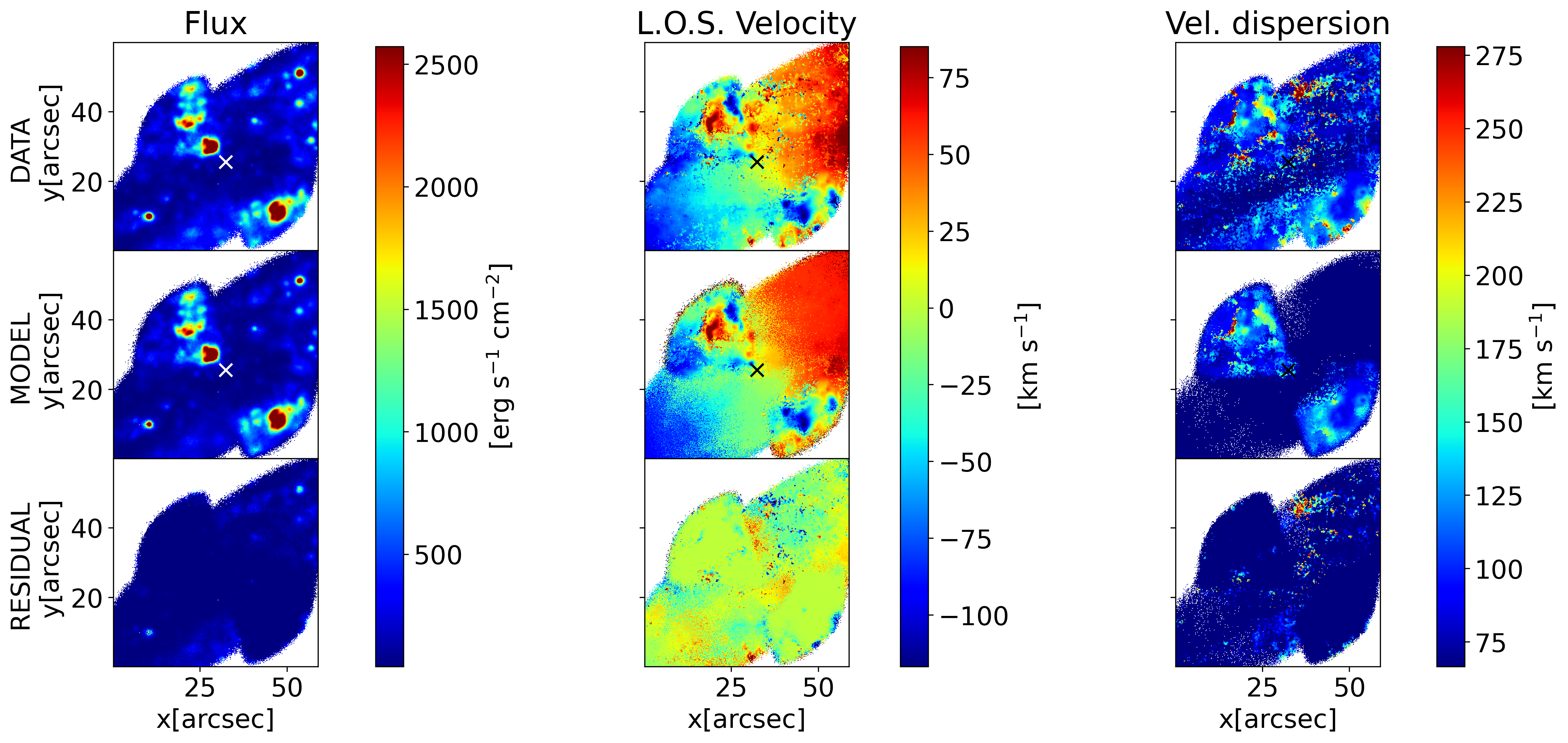}
\caption{Continue of Fig. \ref{fig:weighted_moment_maps_cont1} for NGC 5728, NGC 7582, and Cen A.}\label{fig:weighted_moment_maps_cont2}
\end{figure*}

\begin{figure*}[ht!]
\centering
\includegraphics[width=\textwidth]{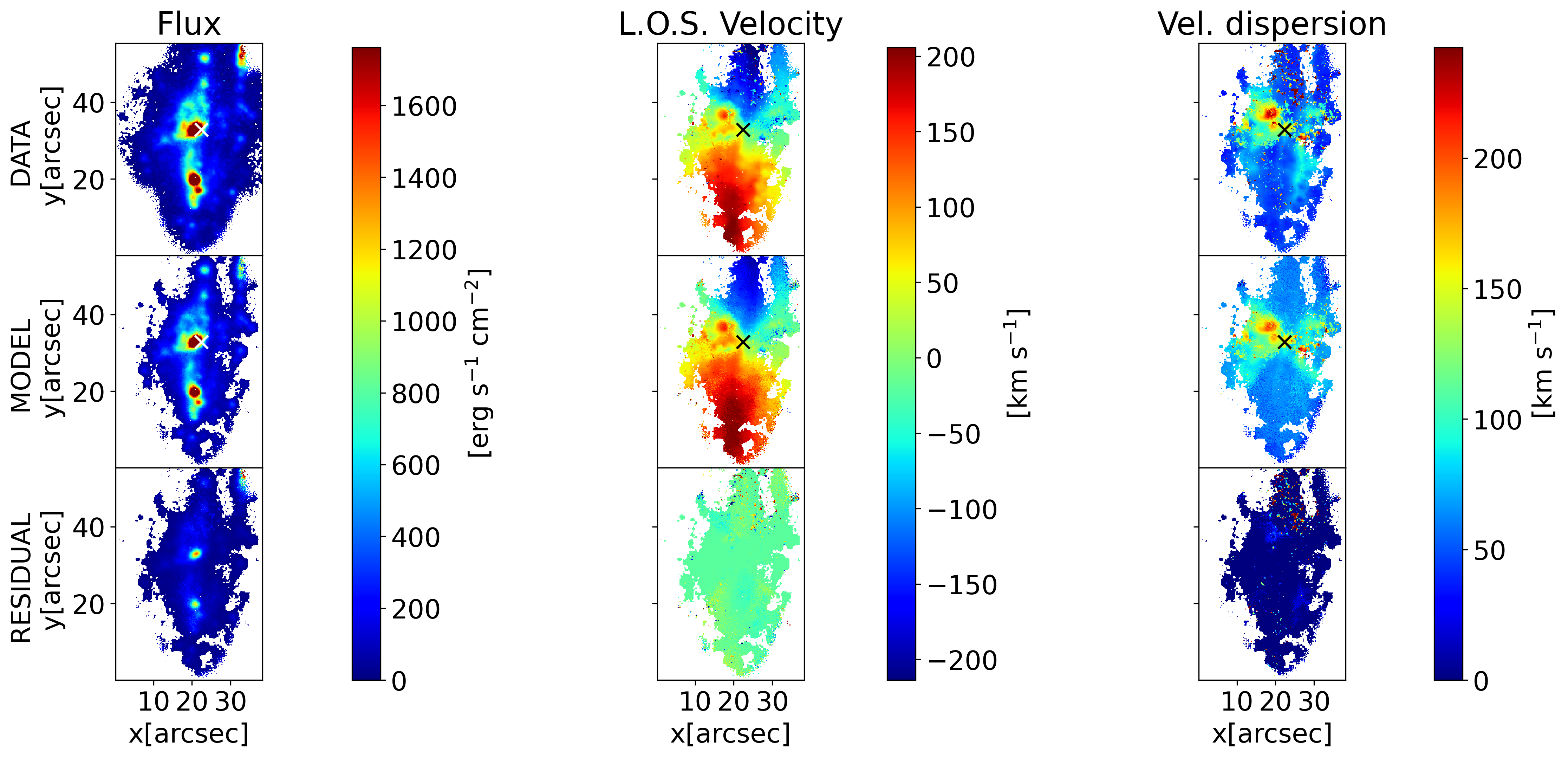}
\includegraphics[width=\textwidth]
{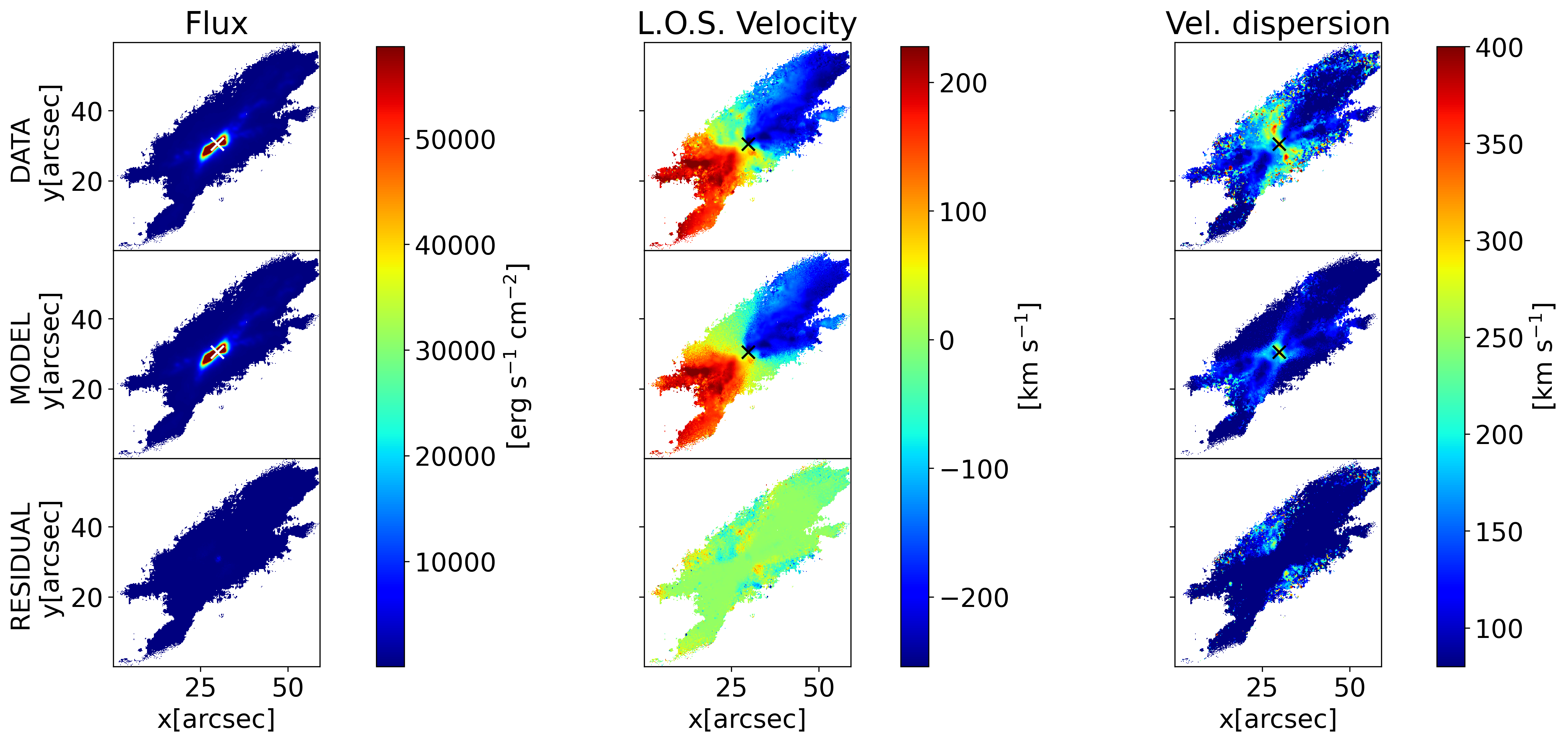}
\caption{Continue of Fig. \ref{fig:weighted_moment_maps_cont1} for IC 1657 and IC 5063.}\label{fig:weighted_moment_maps_cont3}
\end{figure*}


\begin{table*}
\centering
    \addtolength{\leftskip} {-2cm}
    \addtolength{\rightskip}{-2cm}
    \begin{tabular*}{1.2\textwidth}{cc|ccccccccc}
    \hline
    \hline
     \multicolumn{2}{c|}{Galaxy} & \multicolumn{8}{c}{Fit parameters} \\
    \hline
        ID   & $R_{\rm 0}^{*}$ & $<V_{\rm r, blue}>$ & $<V_{\rm r, red}>$ &$\beta_{\rm blue}$ & $\beta_{\rm red}$ &  $\theta^{*}$ & $\gamma^{*}_{\rm blue}$ & $\gamma^{*}_{\rm red}$ & $\beta_{\rm disk}$ & V$^{20 kpc}_{ESC}$ \\
     &  [kpc] & [km s$^{-1}$]  & [km s$^{-1}$]  & [$^{\circ}$]  & [$^{\circ}$]  & [$^{\circ}$] & [$^{\circ}$] & [$^{\circ}$] & [$^{\circ}$] & [km s$^{-1}$] \\
    \hline
    NGC 4945  &  0.54 & 1008 & 930 & 72  & 101  & 100 & 50 & 230 & 80 & 240\\
    Circinus & 0.61 & 515 & --- & 74 &  ---  & 110 & 65 & ---& 61 & 400\\
    NGC 7582  & 3.28 & 650 & 645 & 70 & 112  & 120 & 116 & 296& 80 & 300\\
    Centaurus A   & 0.46 & 760 & 730 & 90  & 93  & 80 & 305 & 125& 107 & 240\\
    NGC 1365 &  2.7 & 975 & 620 & 84  & 92  & 120 & 230 & 50& 125 & 365\\
    NGC 2992 &  5.3 & 700 & 710 & 81  & 120  &  110 & 60 & 240& 80 & 310\\
    IC 5063  & 6.5 & 775 & 1110 & 70  & 110  & 80 & 60 & 230& 80 & 380\\
    NGC 5643   & 2.10 & 710 & 770 & 80  & 86  & 100 & 275 & 95& 75 & 190\\
    NGC 5728   & 6.4 & 800 & 890 & 82  & 100  & 90 & 55 & 235& 80 & 370\\
    IC 1657   & 4.8 & 710 & 650 & 86  & 103 & 100 & 94 & 274& 102 & 290\\

    \hline
\end{tabular*}
\caption{Galaxies properties and best-fit parameters of the ionised outflows in the sample. The asterisks indicate the parameters kept fixed during the fitting procedure. From left to right: Source ID, outflow extension in kpc ($R_{\rm 0}$), average blue-shifted and red-shifted outflow radial velocity ($<V_{\rm r, blue}>$, $<V_{\rm r, red}>$), inclination with respect to the LOS of (bi)conical outflows ($\beta_{\rm blue}$, $\beta_{\rm red}$), outer semi-aperture of the conical model ($\theta$), P.A. measured counter-clockwise from north ($\gamma$), and wind velocity at radius $R_{\rm 0}$ to reach the 20 kpc scale (see Sec. \ref{escape_velocity}) Semi-apertures and position angles are fixed before the fit. Outflow radial velocity profile for each source are shown in Fig. \ref{fig:vel_profiles}. For NGC 2992 we adopted two different outflow radii: the red-shifted and blue-shifted cones have an extension of 35'' and 20'', respectively. }\label{tab_appendix2_outflow_proeprties}\end{table*}

\bmhead{Acknowledgements}
CM, GC, AM, GT, FM, FB, EB and GV acknowledge the support of the INAF Large Grant 2022 "The metal circle: a new sharp view of the baryon cycle up to Cosmic Dawn with the latest generation IFU facilities" and of the grant $PRIN-MUR 2020ACSP5K_002$ financed by European Union – Next Generation EU. EDT is supported by the European Research Council (ERC) under grant agreement no. 101040751. SC and GV acknowledge funding from the European Union (ERC, WINGS,101040227).

\section*{Declarations}
\begin{itemize}
\item Data availability: All data are public and available on the ESO archive processed data portal (\url{https://archive.eso.org/scienceportal/home})
\item Author contribution: C.M. and A.M. designed and coordinated the work, prepared the figures and drafted the manuscript. All authors have contributed to the analysis and interpretation of the data, the results and to the final text.
\end{itemize}


\bigskip





\begin{appendices}
\section{Outflow and Disc inclination}\label{appendixbo_outflow_disc_inclination}
\begin{figure}[h!]
\centering
\includegraphics[width=\textwidth]{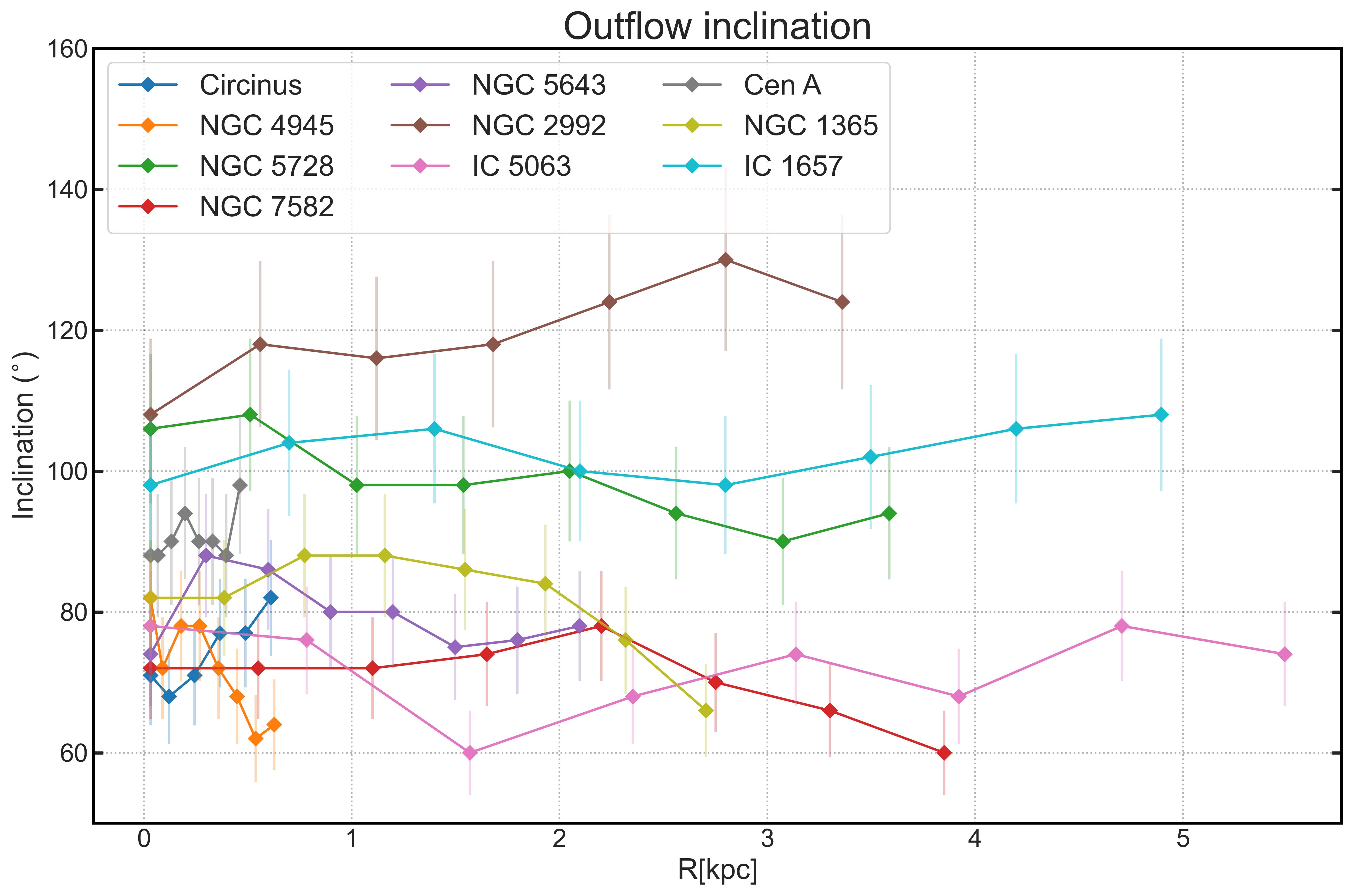}
\includegraphics[width=\textwidth]{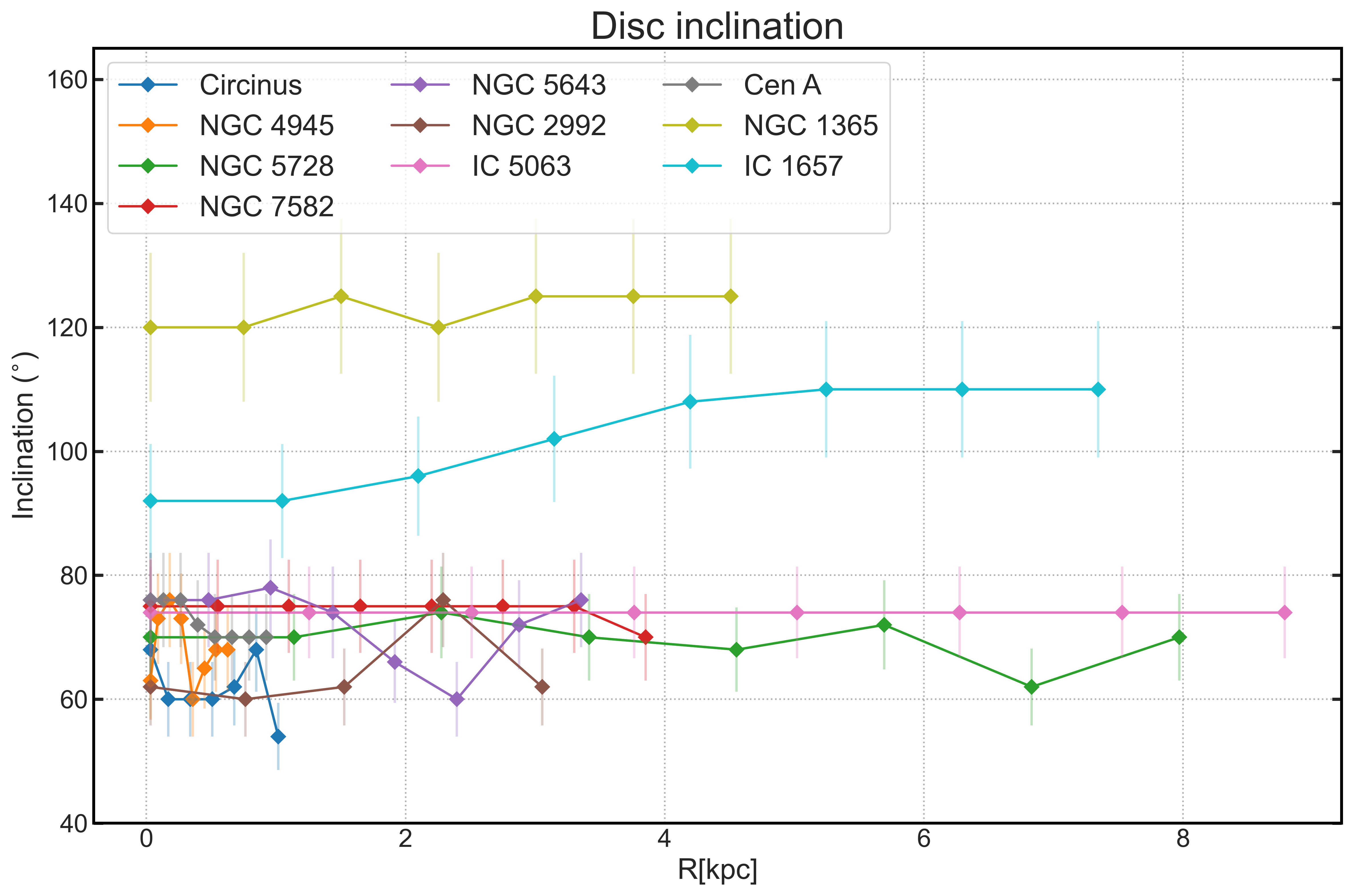}
\caption{Shell inclination for the outflow (top) and disc (bottom) kinematic model as a function of the distance from the AGN in our sample. Uncertainties are computed as described in Sec. \ref{moka_section}. }\label{fig:outflow_disk_inclination}
\end{figure}

As discussed in previous sections, we performed the shell by shell fit of the outflow and disc emission with \MOKA and retrieved for each shell the intrinsic velocity and inclination with respect the line of sight. Figure \ref{fig:outflow_disk_inclination} shows the result for the outflow (top) and disc (bottom) inclination as a function of the distance from the AGN for our MAGNUM sample. The disc inclination in IC 5063 is fixed to the value of 74$^{\circ}$, as measured by \cite{Morganti1998} from HI observations. Our results show that on average the model inclination in different shells are consistent with the average value reported in Table \ref{tab_appendix2_outflow_proeprties}. Uncertainties of the disc and outflow inclination in each shell are obtained as described in the Methods section. The average best fit disc inclination for each source in our sample is consistent with estimates from previous works \citep{Fischer2013, Gaspar2022, lopezcoba2020,Risaliti2013, durre_mould_2018, Ma2023, Morganti2015, garcia_benete_2021}.

\section{Theoretical predictions on AGN outflow acceleration}\label{appendix3_theoretical_prediction}
\begin{figure}[t!]
\centering
\includegraphics[width=\textwidth]{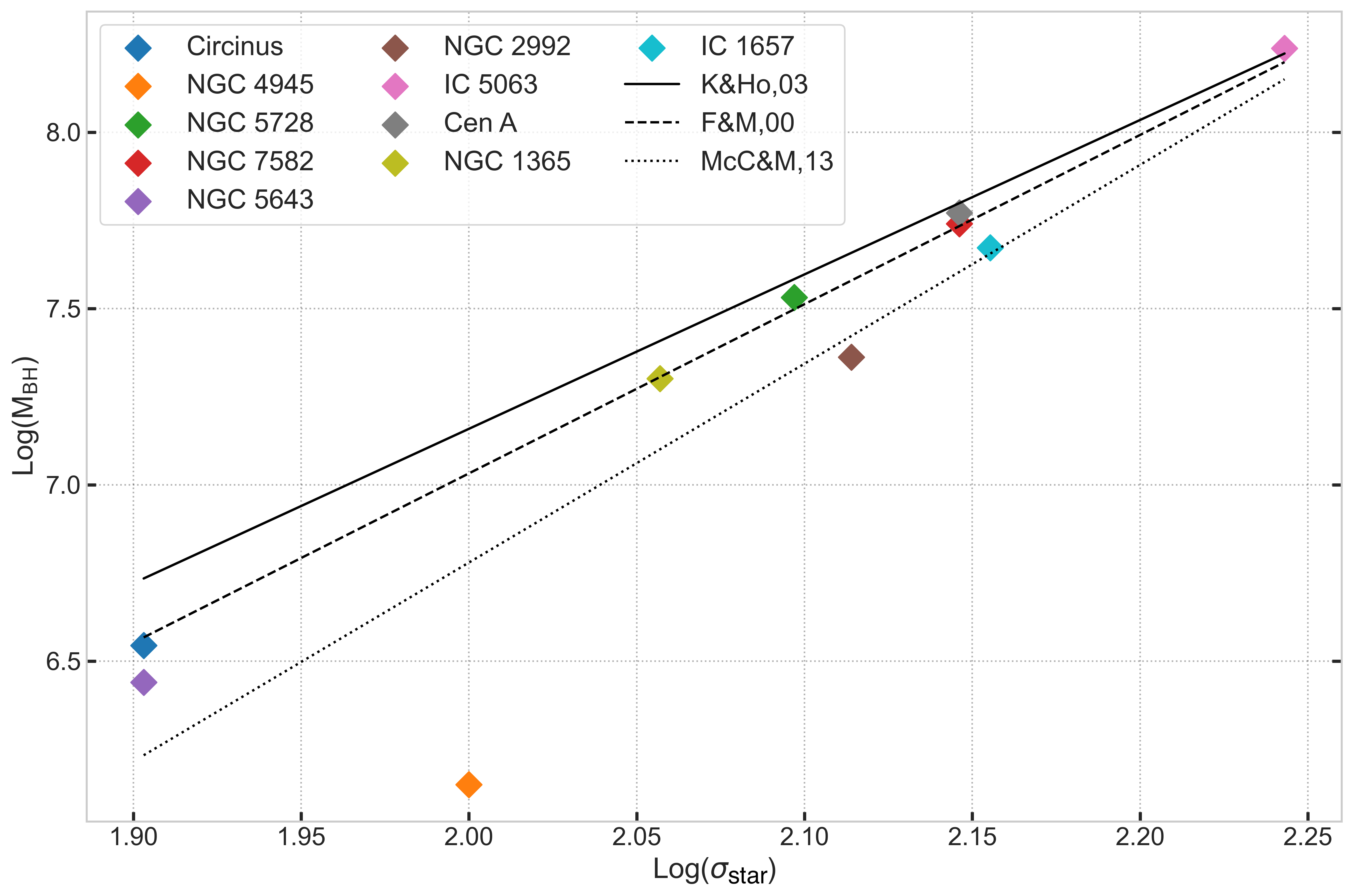}
\caption{Black hole mass vs stellar velocity dispersion relation in our sample. Solid, dashed, and dotted lines represent the expected M-$\sigma$ relations from \cite{Kormendy2003}, \cite{Ferrarese2000}, \cite{McConnell2013}, respectively.}\label{fig:Msig_relation}
\end{figure}
Many theoretical models and hydrodynamical simulations have been proposed to establish the evolution of an expanding gas shell sustained by the AGN radiation pressure. Such models investigate the kinematic and morphology of the gas entrained in the outflow, as a function of gas physical properties and different initial conditions \citep[e.g.][]{Zubovas2012, Zubovas2014, Nayakshin2014, king_pounds_2003, King2010, Thompson2015, Costa2020}.
Among the most credited, \cite{King_pounds_2015} modelled the propagation of a nuclear-scale ultra fast outflow (UFO) as a quasi spherical wind emerging from the SMBH accretion disk. They found that while the wind impacts the ISM, shocks are created and the ISM heats up to T$\sim 10^{11}$ K. These temperatures favour inverse Compton scattering of photons in the AGN radiation field as the main cooling process. In particular, if the shock caused by the wind is close to the SMBH, the radiation field is so intense that the post-shock medium rapidly cools and looses energy to radiation on timescales shorter than its flow time, leading to a momentum-driven regime \citep[see][]{ciotti1997}.

\cite{King_pounds_2015} showed that Compton cooling is efficient up to a critical radius of $\sim$ 1 kpc, whereas outside of it the wind shock energy is trapped and transferred to the outflowing gas (see Eq. 33 in \citep{King_pounds_2015}). This critical radius marks the separation between the inner momentum-conserving wind and the outer energy-conserving wind phases.
In particular, the momentum-conserving regime is established on smaller scales as it is favoured by the intense radiation field of the AGN accretion disk. Instead, during the energy-conserving phase, also referred to as adiabatic phase, the outflow is expected to have enough energy to remove the interstellar gas, suppressing SMBH accretion and quenching star formation by depleting the gas reservoir (see Sec. 5.3 in \cite{King_pounds_2015}). 

\cite{King_pounds_2015} showed that the energy-conserving regime is established when the SMBH mass ($\rm M_{\rm BH}$) reaches its $\rm M_{\sigma}$ value, i.e. $\rm M_{\sigma} \sim 3 \times 10^8 M_{\odot} \sigma^4_{200}$, where $\sigma_{200}$ is the bulge velocity dispersion in units of 200 km s$^{-1}$ \cite[see also][]{power2011}. This model predicts that during this regime the outflow velocity (\rm $v_{\rm ed}$) only depends on the AGN bolometric luminosity (L$_{\rm bol}$) and $\rm M_{B\rm H}$, as: $v_{\rm ed} \sim k \ l^{1/3} \ \sigma^{2/3}_{200} \ \rm km \ s^{-1}$, where $\rm k \sim 10^3$ and $ l = \rm L_{\rm bol}/L_{\rm EDD}$ is the Eddington ratio. Interestingly, the weak dependence of $v_{\rm ed}$ on $l$ tell us that we can observe high velocity outflows on large scales even in a weakly luminous AGN. As previously discussed, we do not expect to witness the transition between the two kinematic regimes precisely at the scales predicted by the mentioned equations due to the extremely simplified assumptions the model rely on. Nevertheless, the clear resemblances between the observed and predicted outflow kinematic lead us to assume such model as qualitatively representative of the observed gas properties.

Observations have shown a tight correlation between M$_{\rm BH}$ and the velocity dispersion of the host bulge, leading to the relation $\rm M_{\rm BH} \sim 3.1 \times 10^8 \ M_{\odot} \sigma^{4.4}_{200}$, which is often interpreted as an upper limit for the M$_{BH}$ at a given bulge velocity dispersion \citep[][]{Ferrarese2000, Gebhardt2000, Kormendy2003, McConnell2013}. Theoretically, if $\rm M_{BH} < M_{\sigma}$ there is not sufficient radiation pressure from the accretion disk to push the wind at large radii, and thus a momentum-conserving regime is established. On the other hand, episodic bursts of accretion can push the black hole mass above this limit ($\rm M > M_{\sigma}$), allowing the wind to reach the critical radius and evolve into the adiabatic phase. Under this scenario depicted by \cite{King_pounds_2015}, the M-$\sigma$ marks the boundary at which outflows evolve from momentum- to energy-conserving, with this transition happening at $\sim$ 1 kpc. Figure \ref{fig:Msig_relation} shows the location of our sample on the M-$\sigma$ relation, with the stellar velocity dispersion of our sources measured by fitting the stellar continuum (for details of the stellar continuum fitting methodology see \cite{tozzi2021, mingozzi2019}). Interestingly, all our sources except NGC 4945 lie on the M-$\sigma$ relation predicted by different models \citep[][]{Kormendy2003, Ferrarese2000, McConnell2013}. 

Theoretical models predict that if the gas accretion onto the SMBH stops, the propagation of the shock driving the wind will be sustained by the residual gas pressure for a time an order of magnitude longer than the duration of the accretion phase (see Fig. 9 in \cite{King_pounds_2015}). In this case, the wind cannot reach the galaxy outskirts and thus clear the galaxy gas content \citep[see][]{Silk2010, power2011}. In \cite{King2011} instead, the authors show that to remove a significant amount of mass it is necessary for the shock to reach the galaxy edge. For this to happen, the black hole needs an intense growth activity with a duration of the order of the Salpeter time (i.e., t$_S \sim 10^7$ yr). The AGN accretion luminosity must be communicated to the host gas and this happens at the adiabatic phase, during which the shock wind energy, which derives from the accretion luminosity, is transferred to the ambient gas which gets trapped into the outflowing material. Therefore, only winds in the energy-conserving regime are energetic enough to reach the galaxy outskirt and clear the host of its gas reservoir \citep[e.g.][]{Murray2005, murray2011}. This is possible only for long-lasting accretion episodes or frequent and intense accretion burst.

\end{appendices}

\bibliography{example}

\end{document}